# Phonon-assisted tunneling in Jahn-Teller E×e impurity centers in crystals

V. Hizhnyakov

*Institute of Physics, University of Tartu, W. Ostwald Street 1, 50411 Tartu, Estonia*

**Abstract**

Tunnel transitions between various distorted states of impurity centers in crystals with the E×e Jahn-Teller effect are considered, taking into account changes in the dynamics of phonons caused by this effect. Both linear and quadratic vibrational interactions are taken into account. It was found that phonon scattering accompanying tunneling leads not only to a broadening of the energy spectrum of the transitions, but also to a deterioration in resonance. The results obtained are consistent with measurements of ultrasonic attenuation in $Al_2O_3$:Ni, GaAs:Mn and in GaAs:Cu doped crystals. A notable feature of E×e tunneling is the existence of a range of values for quadratic vibration interaction, in which the tunneling remains coherent at sufficiently high temperatures.

## 1. Introduction

As shown by Jahn and Teller [1], symmetrical configurations of molecules are unstable in degenerate electronic states due to the interaction of valence electrons with vibrations of the molecule (vibronic interaction). Due to this instability the potential energy surfaces (PES) of molecules acquire several energetically equivalent minima of low symmetry separated by potential barriers. The states of the molecule in these minima, although they may be long-lived, usually are not completely stable: tunnel transitions between these states are possible [2-5]. At zero temperature, these transitions occur as coherent tunnel transitions and they can be considered as pseudo-rotations [3]. This holds also for impurity centers in crystals in degenerate electronic states [6-12]. However, in crystals, due to the presence of phonon continuum, tunnel transitions are accompanied by vibrational relaxation. As a result, as temperature increases, coherent tunnel motion is replaced by incoherent phonon-assisted tunnel transitions between the PES minima. Experimentally, these transitions can be manifested themselves as the phase relaxation, leading to temperature broadening of the zero phonon lines in optical spectra [13], attenuation of photon echo, and ultrasonic attenuation [14].

This paper discusses tunnel transitions in the trigonal center of impurities in a double degenerate electronic state of the $E$ representation interacting with doubly degenerate vibrations of the $e$-representation. The Jahn-Teller effect in this case is a well-known archetype of the Jahn-Teller problem, called as the E×e problem. If we consider only the linear vibrational interaction, then PES has axial symmetry with a conical intersection and an equilateral circle of minima (the shape of PES resembles a Mexican hat). However, the quadratic vibration interaction leads to the violation of axial symmetry and the formation of three energy-equivalent minima in the PES. The zero-point states in these minima are inherently unstable with respect to the tunnel transitions under the barrier. As a result, the ground state is divided into one double degenerate tunnel state and one fully symmetric tunnel state [2]. Usually, a doubly degenerate state is a ground state; However, the reverse order of the tunnel state is possible if the quadratic vibronic interaction dominates [15,16]. Splitting is highly dependent on the strength of the vibronic interaction, rapidly decreasing as it increases (see Figs. 2 - 5 in [16], which present the results of calculating the energies of lower levels of the E×e state for different linear and quadratic vibronic interactions). In crystals, the upper tunnel state slowly decays with the emission of phonon [17,18]. At higher temperatures, two-phonon relaxation processes of Raman type become significant, leading to the replacement of coherent tunneling with incoherent tunneling.

Here we focus on incoherent tunnel transitions and pay attention to the case of a strong Jan-Thieler effect. In this case, there is a strong mixing of phonon continuum (strong Duschinsky effect) during transitions. Until now, this case has remained unsolved due to mathematical difficulties. To solve it, we use the resently developed theory of electronic transitions [19,20], which is true for arbitrary linear and quadratic vibronic interactions. Using this theory, it was possible to calculate the tunnel transition rate, taking into account, within the framework of the non-perturbative theory, two-phonon processes of the Raman type. It has been established that these processes lead not only to a broadening of the spectrum of tunnel transitions, but also to a change (increase) in the average frequency of the spectrum of these transitions. At low temperatures $T \ll T_D$, the width and the average frequency of the spectrum of tunnel transitions depend on temperature as $\gamma \propto T^7/T_D{}^7$ and $\delta \propto T^4/T_D{}^4$, respectively, where $T_D$ is the Debye temperature. As a result, the Raman type processes cause not only a weakening, but also deterioration of the resonance tuning. The reason for the change in average frequency lies in the zero-point vibrations, which make the creation of a phonon during the Raman process more likely than the destruction of a phonon of close frequency. In perfect crystals with single centers, the rate of incoherent tunnel transitions depends on $\gamma$ and $\delta$ according to the relation $\Omega \propto \gamma/(\delta^2 + \gamma^2)$; at low temperatures it is inversely temperature-related: $\Omega \propto T^{-1}$. These results also apply to tunnel transitions between PES minima in other Jan-Teller systems. Note that the Raman relaxation mechanism in tunnel transitions has also been considered in [21] on the basis of perturbation theory. However, the greatest effect - an increase in the average frequency of transitions with temperature - was not taken into account.

The temperature dependences of the width and average frequency of the spectrum of tunnel transitions are similar to these dependences the zero-phone line (ZPL) in the optical spectra of impurities in crystals [22,23]. However, the reasons for the dependence of the average frequency on temperature at tunnel transitions and ZPL are different: this frequency ZPL is mainly dependent on the first-order quadratic vibration interaction [24-26], but in tunnel transitions it is determined by Raman processes described by the second-order quadratic vibrational interaction (the effect of the first-order quadratic interaction on tunneling in Jahn-Teller systems is cancelled out by symmetry arguments).

A notable phenomenon we found is that in the E×e case, there is a range for the values of the quadratic vibronic interaction around the critical value of $b_0 = 1/9$ (for the Debye frequency $\omega_D$ as one), at which the effect of this interaction on the tunnel transitions disappears. As a result, for such quadratic vibronic interaction, tunnel transitions remain coherent at a sufficiently high temperature. The reason is that in the E×e case quadratic vibronic interaction leads to the formation of tunnel barriers. If this interaction is weak, then the barriers are small, and the vibrations along the PES through are soft, being much softer than the vibrations across the through [10,27]. As $b$ increases, so do the barriers. Then the vibrations along the trough become harder, but vibrations across the through are softened. Because of this, the change in phonon dynamics during the tunnel transition first decreases, then disappears at the critical value of the quadratic vibration interaction, and then begins to increase with the increase in this interaction. In the case of the quadratic vibronic interaction close to the critical value, the Raman processes stop, and the tunneling remains coherent at sufficiently high temperatures.

With an increase in temperature, transitions between different PES minimums with creation and destruction of several phonons became important, which leads to the replacement of a decrease in the rate of transitions with an increase in it with an increase in temperature [10-12,14]. At high temperatures, the transitions turn into the classical barrier jumps, described by Arrhenius' law [12]. Thus, the change in the tunneling rate from decreasing to increasing with decreasing temperature indicates the emergence of quantum tunneling. In experiments on ultrasound attenuation by $E \times e$ impurity centers of $Ni^{2+}$ ions in $Al_2O_3$ crystals, such a change in the temperature dependences of the transition rate between different PES minima was actually observed at 2 K [14]. A similar effect was observed in the ultrasound attenuation due to tunnel transitions in the $T \times t$ Jahn-Teller centers with tetragonal distortion in doped GaAs:Mn [28] and GaAs:Cu [29] crystals. However, the theoretical explanation for these changes of decrease to increase in $\Omega$ with $T$ was not found in these works.

2. General

The vibronic Hamiltonian of the impurity center in the crystal in the E×e case is [9-11, 16-18]

$$H = H_0 \cdot I + V \quad (1)$$

Here $H_0 = (1/2)\sum_{\iota j} (\omega_j^2 x_{\iota j}^2 - \partial^2/\partial x_{\iota j}^2)$ is the Hamiltonian of phonons of the crystal with the impurity center in the totally symmetric electronic state, $\iota = \eta, \varepsilon$ is the number of line of the $e$ representation, $j = 1, 2, \ldots N$ is the phonon number, $\omega_j$ is frequency and $x_{\iota j}$ is the normal coordinate of a $e$-phonon, $N \sim N_A$, $N_A$ is the Avogadro's number, $I$ is the $2 \times 2$ unit matrix, $V$ is the $2 \times 2$ vibronic interaction mastrix, $\hbar = 1$. Using the electronic functions $|\vartheta\rangle$ and $|\varepsilon\rangle$ of $E$ representation, transforming as $3z^2 - r^2$ and $x^2 - y^2$, respectively, $V$ gets the form [9-11]

$$V = a\left(-Q_\eta \sigma_x + Q_\varepsilon \sigma_z\right) + b\left((Q_\varepsilon^2 - Q_\eta^2)\sigma_z + 2Q_\nu Q_\varepsilon \sigma_x\right). \quad (2)$$

Here $\sigma_x$ and $\sigma_z$ are Pauli matrices, $a$ and $b$ are the parameters of linear and quadratic vibronic interaction, respectively, $Q_\eta$ and $Q_\varepsilon$ are symmetrized configurational coordinates, which are linear combinations of reduced displacements $r_L^\alpha = \sqrt{M_L} R_L^\alpha$ of a small number of close atoms (ions) to the impurity atom (ion), $L$ is the number of the atom (ion) in the center and in the vicinity, $R_L^\alpha$ is the $\alpha = x, y, z$ Cartesian component of the coordinate of the atom (ion) number $L$ and of the mass $M_L$. In a trigonal molecule, the configurational coordinates $Q_\eta$ and $Q_\varepsilon$ are linear combinations of several normal coordinates. In contrast, in the case of an impurity trigonal molecule (center) in a crystal, $Q_\eta$ and $Q_\varepsilon$ are linear combinations of a huge number ~$N_A$ normal coordinates of the entire crystal, which are of the form [16-18] $Q_{\eta,\varepsilon} = \sum_j \varsigma_{\eta,\varepsilon j} x_{\eta,\varepsilon j}$. Here $\varsigma_{\eta j}$ and $\varsigma_{\varepsilon j}$ , are the unite polarization vectors of phonons that satisfy the orthogonality condition $\sum_j e_{\nu j} e_{\nu' j} = \delta_{\nu,\nu'}$, $\nu, \nu' = \eta, \varepsilon$. The density of the states (DOS) of the phonons contributing to the configurational coordinates $Q_\eta$ and $Q_\varepsilon$ is $\rho_0(\omega) = \sum_j \left|\varsigma_{\eta j}\right|^2 \delta(\omega - \omega_j)$; here $\left|\varsigma_{\eta j}\right| = \left|\varsigma_{\varepsilon j}\right|$, $\omega$ is the phonon frequency. For a picture of displacements $Q_\eta$ and $Q_\varepsilon$, see, for example, Fig. 3.7 in [10].

Let us introduce the variables $Q = \sqrt{Q_\eta^2 + Q_\varepsilon^2}$ and $\varphi = \arcsin\left(Q_\eta/Q\right)$ and consider the electronic states $\left|\psi_\varphi\right\rangle = \cos\varphi\left|\vartheta\right\rangle + \sin\varphi\left|\varepsilon\right\rangle$, $\left|\psi'_\varphi\right\rangle = -\sin\varphi\left|\vartheta\right\rangle + \cos\varphi\left|\varepsilon\right\rangle$. These states bring the matrix of vibronic interaction into a diagonal shape with the diagonal elements [10]

$$V = \pm aQ\sqrt{1 + (b/a)Q\cos(3\varphi) + (b/a)^2 Q^2}\,. \quad (3)$$

There are three values of $\varphi$ angle, which correspond to PES minima [10]. In the case $a, b > 0$, they are $\varphi_l = 2\pi(l-1))/3$, where $l = 1, 2, 3$ is the number of one of the PES minima.

Consider small vibrations near the PES minima, and consider the case $3(bQ_0/2a)^2 \ll 1$, where $Q_0$ is the value of $Q$ at the minima. (For $b \le 0.2$, the neglected term is at least 40 times smaller than the counted one; for the highest possible value $b = 0.5$ for PES with one conical intersection, it is 16 times smaller.) Then, near the PES minimum number $l$, $V_l \approx -aQ - (bQ^2/2)\cos(3\varphi_l)$, where the ratios $Q \cong Q_{1l} + Q_{2l}^2/(2Q_0)$, $Q^2\cos(3\varphi) \cong Q_{1;l}{}^2 - 9Q_{2;l}^2/2$ are valid. Here $Q_{\iota l}$ stand for the displacements along and across the direction of the PES minimum number $l$ to the origin ( $Q_{1l} = \sum_j (\varsigma_{\eta j} \cos\varphi_l x_{\varepsilon j} + \varsigma_{\varepsilon j} \sin\varphi_l x_{\eta j})$, $Q_{2;l} = \sum_j (\varsigma_{\varepsilon j} \cos\varphi_l x_{\varepsilon j} - \varsigma_{\eta j} \sin\varphi_l x_{\eta j}))$. Thus, for $b \le 0.5$, the Hamiltonian of small vibrations near the minimum number $l$ of PES, can be represented as

$$H_l \cong H_0 - aQ_{1l} + b_1 Q_{1l}^2/2 + b_2 Q_{2l}^2/2, \tag{4}$$

where $b_1 = -b$, Q. The linear term in the Hamiltonian (4) is responsible for Jahn-Teller distortion, resulting in PES minimum with reduced symmetry of the atomic configuration; $D_{3h} \to C_{2v}$. The configurational coordinates $Q_1$ and $Q_2$ in the minimum number $l$ have the following values: $Q_{1l} = Q_0$, $Q_{2l} = 0$. Here $Q_0 \equiv \sum_j (\varsigma_{\eta j}\cos\varphi_l x_{0\eta j} + \varsigma_{\varepsilon j}\sin\varphi_l x_{0\varepsilon j})$. Simple calculations give $Q_0 = \bar{a}\kappa$ $\bar{a} = a/(1 + \kappa b)$, $\kappa = \sum_j \left|\varsigma_{\eta j}/\omega_j\right|^2$, $x_{0\eta,\varepsilon j} = \bar{a}\varsigma_{\eta,\varepsilon j}/\omega_j^2$. Only the equilibrium positions of vibrations along the direction of the minima to the origin change.

Using the shift transformation $H_l \cong e^{\nabla_l}\tilde{H}_l e^{-\nabla_l} - E_{JT}$, the linear term in Eq. (4) can be excluded. Here $\nabla_l = \sum_j \left(x_{0\eta j}\cos\varphi_l\, \partial/\partial x_{\eta j} + x_{0\varepsilon j}\sin\varphi_l\, \partial/\partial x_{\varepsilon j}\right)$, $E_{JT} = \bar{a}^2\kappa/2$ is the Jahn-Teller stabilization energy. We obtain

$$\tilde{H}_l \cong H_0 + b_1 Q_{1l}^2/2 + b_2 Q_{2l}^2/2 \tag{5}$$

This Hamiltonian describes small vibrations at the PES minimum number $l$. Here, the $\propto Q_1^2$ term describes the vibronic interaction of vibrations along the direction to origin from the minimum number $l$, and the $\propto Q_2^2$ term describes this interaction for vibrations across the direction to origin. The strengths of these interactions $b_1$ and $b_2$ vary (an exception is the case $b = b_0 = 2/(9\kappa)$, when $b_1 = b_2$). Due to these terms, the vibrations along and across the directions to the origin given by the Hamiltonian (5) have different spectrum. In addition, these vibrations have different orientation for different PES minima. Especially big differences are observed in the case of small $b$, when PES looks like a slightly deformed Mexican hat with deep trough and three shallow minima. In this case, the spectrum of vibrations occurring along the direction to the origin (across the trough) practically does not change, but the vibrations occurring along the trough (across the direction to the origin) became soft [26], with small average frequency of order $\omega_0 \sim \sqrt{b}$ and large amplitude of their zero-point vibrations being comparable to $Q_0$. With an increase in $b$, the difference $b_1 - b_2$ in the quadratic vibronic interaction for both vibrations decreases, and in the critical value $b = b_0$ it disappears giving $\tilde{H}_l = \tilde{H}_{l'}.(l, l' = 1, 2, 3)$. At this point, the low symmetry of configurations of atoms in the minima is restored to a higher symmetry; $C_{2v} \to D_{3h}$. For this critical value of $b$, the phonon Hamiltonians in different PES minima have different equilibrium positions, but the same spectra of phonos. As $b$ increases, the $Q_1$ vibrations become softer than the $Q_2$ vibrations, and the symmetry of the atomic configurations in the PES minimuma is restored to reduced symmetry; $D_{3h} \to C_{2v}$.

It should be emphasized that the equation (5) for Hamiltonians of vibrations near PES minima is valid for the dominant linear vibronic interaction when $Q_0 \gg 1$ and $b < 1/2$. In this case, the PES has three minima and one conical intersection (for large $b$ values, this may not be the case: three additional conical intersections can come to the one of the center [15]) and the amplitude of zero-point vibrations $x_0 \sim 1/\sqrt{2\omega_0}$ is small as compared to the distance between the minima). Corresponding condition reads $b \gg (\hbar\omega_D/12E_{jT})^2$. If we assume $E_{JT}/\hbar\omega_D \sim 5$, then $b > 3 \cdot 10^{-4}$.

### 3. Phonon dynamics perturbed by Jahn-Teller effect

To quantitatively characterize the change in the phonon dynamics caused by the Jahn-Teller effect, we use the phonon Green's functions method [30]. According to this method, a change of the single elastic spring by $b_i$ causes a change in the phonon Green's function described by the equation $G_i(\omega) = G(\omega)/(1 - b_i G(\omega))$, where

$$G(\omega) = \int d\omega' \, \rho_0(\omega')/\left(\omega^2 - (\omega' - i0_+)^2\right) \tag{6}$$

is the Green function of unperturbed phonons contributing to the corresponding elastic spring (here and below integrals are made from $-\infty$ to $\infty$, unless otherwise noted). The quantum tunneling occurs at low temperatures, when only low-frequency acoustic phonons are present in the crystal. To take these phonons into account, we use the Debye-Van Hove model. In this model the DOS of phonons, contributing to the reduced displacements, equals to $\rho_0(\omega) = 8\pi^{-1}\omega^2\sqrt{1-\omega^2}$ ($0 \le \omega \le 1$, the Debye frequency $\omega_D$ is taken as one). Here the factor $\omega^2$ takes into account the density of phonons in the Debye model; the factor $\sqrt{1-\omega^2}$ takes into account the Van Hove singularity for the maximum wave vector of phonons. We keep in mind that only even phonon waves contribute to $Q_i$. These waves have zero amplitude in the central atom and $\propto \omega$ amplitude in the neighboring atoms. Therefore, the DOS of phonons involved in tunnel transitions has an additional multiplier $\omega^2$, which gives $\rho(\omega) = 32\pi^{-1}\omega^4\sqrt{1-\omega^2}$ ($0 \le \omega \le 1$). (Note, that in [17,18] this DOS was used to describe the phonon-induced relaxation of E×e Jahn-Teller systems; in [20] it was used to find the characteristics of the ZPL in the optical spectra of impurity centers in crystals.) The Green's function of phonons contributing to the corresponding elastis spring equals

$$\begin{aligned} \mathrm{Re}G(\omega) &\equiv \Delta(\omega) = -2 - 8\omega^2 + 16\omega^3\left(\omega - \sqrt{\omega^2 - 1}\Theta(\omega - 1)\right), \\ \mathrm{Im}G(\omega) &\equiv \Gamma(\omega) = (16/\pi)\omega^3\sqrt{1-\omega^2}, \quad 0 \le \omega \le 1. \end{aligned} \tag{7}$$

In this model $\kappa = -G(0) = 2$ and $b_0 = 1/9$. The spectra of perturbed vibrations $Q_{1l}$ and $Q_{2l}$ in PES mina participating in tunnel transitions are equal to

$$\Gamma_{1,2}(\omega) = \Gamma(\omega)/\left((1 - b_{1,2}\Delta(\omega))^2 + b_{1,2}^2\Gamma^2(\omega)\right). \tag{8}$$

The results of the calculation of these spectra are shown in Fig. 1.

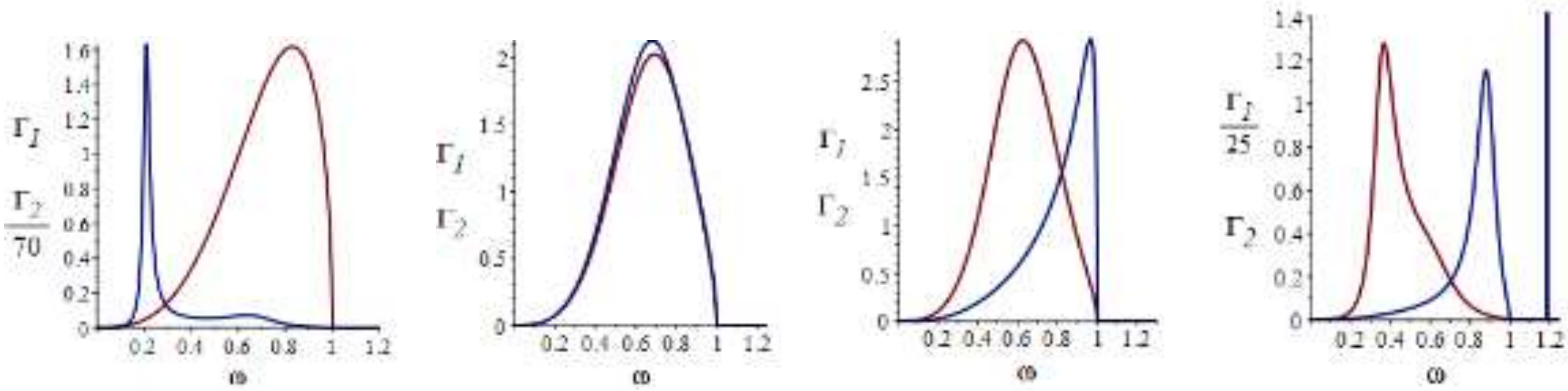

Fig. 1. Spectra of phonons contributing to E×e tunnel transitions. $\Gamma_1$ (brown line) and $\Gamma_2$ (blue line) are the DOS of vibrations along and across the direction to the origin, respectively. Left figure corresponds to quadratic vibronic interaction $b = 0.02$, left middle to $b = 0.109$, right middle to $b = 1/6$ and right to $b = 0.35$.

The case of almost linear vibronic interaction corresponds to a significant $a$, small $b_1 \cong -b$, and negative $b_2 \approx -1/\kappa = -1/2$, when the shape of PES has three shallow minima. In this case, spectra $\Gamma_1(\omega)$ and $\Gamma_2(\omega)$ are very different: the frequencies of vibrations occurring across the direction to the origin (along the through) are significantly less than the frequencies of vibrations occurring along the direction to the origin. In the case of very small $b$, a pseudolocal vibration of low frequency $\omega_0 \cong 1.5\sqrt{b}$ is manifested in the spectrum of $Q_2$ vibrations occurring across the direction to the origin (along the trough; see Fig. 1, left). With increasing $b$, the difference in $|b_1 - b_2|$ and in phonon spectra across and along the direction to the origin decreases, and for $b$ close to $b_0 = 2/(9\kappa) = 1/9$ it disappears (see Fig. 1, left middle). In this case $\tilde{H}_l = \tilde{H}_{l'}$. For this critical value of $b$, the phonon Hamiltonians in different PES minima have different equilibrium positions, but the same spectra of phonons. With larger $b$, $Q_1$ vibrations become softer than $Q_2$ vibrations (see Fig. 1, right middle). For $b > 1/6$, a local mode appears in phonon spectrum for for $Q_2$ vibrations with the frequency above the phonon spectrum (see Fig. 1, right). For $b$ close to $\kappa^{-1}$ (but less than $\kappa^{-1}$), in addition to the hard local $Q_2$ mode, a low frequency pseudolocal $Q_1$ mode also appears in the phonon spectrum. A strong difference of vibrations at PES minima for large $|b_1 - b_2|$ leads to a strong phonon mixing during a tunnel transition (strong Duschinsky's rotation; see also [19,20]. where strong Duschinsky's rotation is examined in more detail). However, for $b$ close to $b_0 = 1/9$, which corresponds to small $|b_1 - b_2|$, the Duschnsky rotation of phonons during the tunnel transitions disappears.

## 4. Rate of tunnel transitions

We consider tunnel transitions in the case when the Jan-Teller system is most of the time in one of the minima and rarely and randomly makes transitions from one minimum to another. Due to vibronic interactions, these transitions are accompanied by Raman-type processes with creation and destruction of phonons of close energy (frequency). In this case, Raman processes cause a broadening of the transition spectrum and a change in the average frequency of this spectrum. Both these effect determine the rate of transitions. To find this rate, we apply Fermi's golden rule. We get the following equation for this rate [10]: $\Omega = 2\pi \left\langle \sum_f \left| H_{ll';nf} \right|^2 \delta(E_{1n} - E_{l'f}) \right\rangle$, where $H_{ll';nf} \approx (E_{JT}/2) \left\langle \Psi_n^{(l)} \middle| \Psi_f^{(l')} \right\rangle$ is the non-diagonal matrix element of the Hamiltonian for the initial ($\left| \Psi_n^{(l)} \right\rangle$) and final ($\left| \Psi_f^{(l')} \right\rangle$) phonon state in the minimums number $l$ and $l'$, respectively. Using the integral representation for the delta-function $\delta(x) = \int e^{ixt} (dt/2\pi)$ and the Schrödinger equation, the transition rate can be presented as

$$\Omega \approx (E_{JT}/2)^2 \int dt F(t), \tag{9}$$

where $F(t) = \left\langle e^{itH_{l'}} e^{-itH_l} \right\rangle$ is the Fourier transform of the energy (frequency) spectrum of the tunnel transitions, $\langle \ldots \rangle$ is the quantum-statistical averaging over the initial vibrational states [31]. The equation (9) to the constant factor is analogous to the equation for the optical spectrum of an impurity center [23,31]; the difference is that in the case of tunneling, $H_l$ and $H_{l'}$ describe vibrations in energy equivalent states, while in the case of optical transition there is no such equivalence. Note, that equation (9) is also true for tunnel transitions in other Jan-Teller systems, including the case of T×t. In perfect crystals, the minima have the same energy. In real crystals, due to presence of the inhomogeneous field resulting from crystal defects, the PES minima have a slightly different energy. The influence of this field can be taken into account by adding the factor $e^{-\gamma_0|t|}$ in the Fourier integral in equation Eq. (9), where $\gamma_0$ is the width of the distribution of the energy differences in the PES minima.

At low temperatures, the largest contribution to $\Omega$ comes from zero-phonon transitions, having small spectral width [23,24]. This contribution is given by $F(t)$ in the large $t$ limit. Usually, $F(t)$ is calculated using a cumulant decomposition with respect to the quadratic vibronic interaction and taking into account up to the second order terms [23,24]. This is only correct if the quadratic vibronic interaction is weak. However, the asymptotic value $F(t)_{t\to\infty}$ can also be found for the arbitrary strength of the quadratic vibronic interaction. This was first done in [32] by summing cumulants if only one configurational coordinate is involved in this interaction. A more detail description of this case can be found in Refs. [33,34] (in Ref. [35], the ZPL width in a special case case of two configurational coordinates was also considered). However, in the case of tunnel transitions under consideration, there are at least two pairs of configuration coordinates describing vibrations along and across the direction to the origin at each PES minimum that contribute to tunnel transitions. To find $F(t)_{t\to\infty}$ in the case of multiple configurational coordinates involved in the quadratic vibronic interaction, we apply a method developed in [19,20], which does not use cumulant decomposition. This method is based on the use of the Stratonovich identity [36] $e^{x^2/2} = (2\pi)^{-1/2}\int du e^{xu-u^2/2}$, which allows us to consider quadratic vibronic interaction as a randomly fluctuating linear vibronic interaction. The latter interaction can be taken into account in the general case [19]. We obtain $F(t)_{t\to\infty} = e^{-(\gamma+i\delta)t-Z}$, where

$$\gamma - i\delta = \frac{1}{4\pi}\int d\omega \ln\left(\left(1-\bar{b}D_1(\omega)\right)\left(1+\frac{\bar{b}}{4}\left(D_1(\omega)+3D_2(\omega)\right)\right)+\frac{\bar{b}^2}{4}D_1^2(\omega)\right), \quad (10)$$

$\bar{b} = b_2 - b_1$, $D_\iota(\omega) = G_\iota(\omega) + in_{|\omega|}\Gamma_\iota(|\omega|)$, $Z = \bar{a}^2\int d\omega\omega^{-2}\Gamma_1(\omega)(2n_\omega+1)$ is the Huang-Rhys factor, $n_\omega = 1/(e^{\omega/k_bT}-1)$ is the Planck population factor (see Appendix). At low temperatures

$$\gamma = \left(34560\zeta(7)/\pi^2\right)B^2\,(T/T_D)^7\,, \quad (11)$$

$$\delta = \left((58b+6)\pi^3/5\right)B^2\,(T/T_D)^4\,, \quad (12)$$

$$Z \approx \left(16/(3\pi(1-2b)^2)\right)(E_{JT}/\omega_D)\left(1+\pi^2(T/T_D)^2\right), \quad (13)$$

where $B = (1-9b)/\left(7b(1-2b)\right)$, $\zeta(7) = 1.008$ ($\zeta(x)$ is zeta-function), Substituting the equation (10) into $F(t)_{t\to\infty}$ and adding inhomogeneous width $\gamma_0$ to $\gamma$, we obtain

$$\Omega \approx (E_{JT}/2)^2\,e^{-Z}(\gamma+\gamma_0)/\left(\delta^2+(\gamma+\gamma_0)^2\right). \quad (14)$$

The results of calculations $\gamma$, $\delta$ and $\Omega$ for the Debye-van Hove model are presented in Fig. 2.

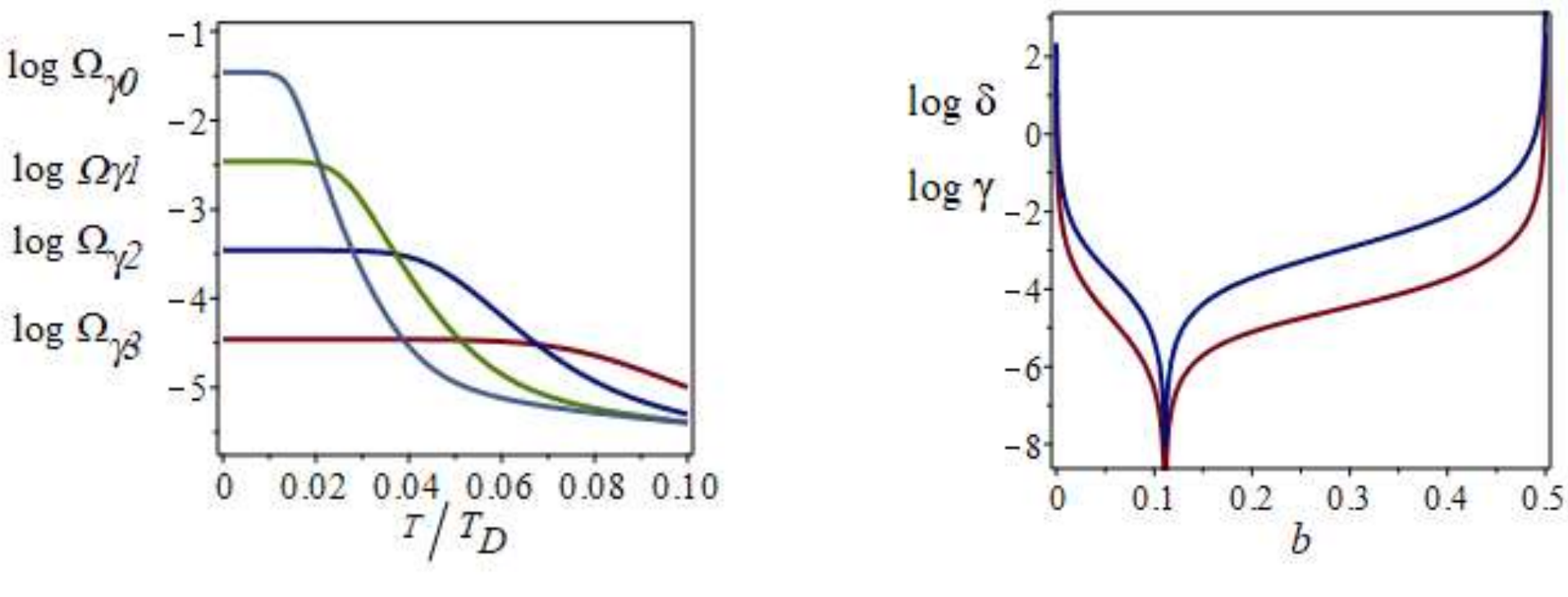

Fig. 2. The rate $(\Omega)$, average frequency ($\delta$) and spectral width ($\gamma$) of tunnel transitions in the Jahn-Teller $E\times e$ impurity center in the crystal. Left: the dependence of $\log_{10}\Omega$ on temperature for $b = 0.05$, Z=7 and for inhomogeneous widths $\gamma0 = 10^{-5}$, $\gamma1 = 10^{-4}$, $\gamma2 = 10^{-3}$, $\gamma3 = 10^{-2}$; right: the dependence of $\log_{10}(\gamma)$ (lower line) and $\log_{10}(\delta)$ (upper line) on the quadratic vibronic interaction $b$ (in $\omega_D$ units).for $T = 0.05T_D$.

## 5. Discussion

The presented equations (11) and (12) describe the dependence of the width and the average frequency of the spectrum of tunnel transitions, caused by the two-phonon (Raman) processes, on the strength of the quadratic vibronic interaction and temperature in the low temperature range. The temperature dependences $\gamma$ and $\delta$ are similar to the temperature dependences of the width and the average frequency ZPL in the optical spectrum of the impurity center at low temperature [22-24]. However, the reasons for the temperature dependence of the average frequency of tunne transitions and ZPL are different: the ZPL average frequency is determined mainly by the quadratic vibronic interaction in the first order [23,24]; but in tunnel transitions it is given by the quadratic vibronic interaction in the second orderer. The reason for the latter change lies in the zero-point vibrations, which makes the creation of a phonon more likely than the destruction of a phonon of the same frequency. The effect of the first-order quadratic vibronic interaction on the average frequency of spectrum tunnel transitions in Jahn-Teller systems is cancelled out by symmetry arguments.

According to (14), the rate of tunnel transitions $\Omega$ decreases with increasing temperature (see also Fig.2, left). If the temperature is so low that $\gamma$ is less than the inhomogeneous width $\gamma_0$, then $\Omega$ decreases with temperature according to the law $\Omega \propto (T^8 + Const)^{-1}$. At high temperatures, multiphonon processes, which were not taken into account here, become important, causing an increase in $\Omega$ with increasing temperature. Finally, these processes turn into classical barrier hops described by Arrhenius' law $\Omega = 2\bar{\omega}_0 e^{-V_0/k_B T}$ [14,28] ($V_0 = bQ_0^2$ is the height of the barrier, $\bar{\omega}_0$ is the average frequency of $Q_2$-vibrations). Thus, the decrease in $\Omega$ with an increase in $T$ is a characteristic feature of quantum tunneling, distinguishing it from the classical hopping processes.

The dependences of the width and the average frequency of tunnel transitions on the quadratic vibronic interaction $b$ in Debye-Van Hove model are presented in Fig. 2 (right). From this figure, it can be seen that $\delta$ exceeds $\gamma$. As a result, in the case of a weak inhomogeneous crystal field at low temperatures, $\Omega \approx (E_{JT}/2)^2 e^{-Z} \gamma/\delta^2$, which gives

$$\Omega \approx \left(7(b + 3/29)B\right)^{-2} e^{-Z} \left(E_{JT}^2/\omega_D\right) T_D/T \tag{15}$$

This means that in this case the rate of the tunnel transitions depends on temperature according to the inverse linear law: $\Omega \propto T^{-1}$. Such a dependence of the tunnel transitions rate on temperature was recorded [28] in experiments for ultrasonic attenuation for Mn impurity centers in a GaAs crystal with the $T \times t$ Jahn-Tellr effect for temperatures $7 \div 10\,\text{K}$ and frequencies $390 \div 1400\,\text{MHz}$. (According to the reference [14,28], ultrasonic attenuation in crystals with Jahn-Telller impurity centers peaks at the frequency $\omega \sim \Omega$.) At higher temperature, the decrease in $\Omega$ with temperature is replace by an increase. A similar change in the increase to decrease in the rate of tunnel transitions with increasing temperature was also observed for ultrasonic attenuation in $T \times t$ Cu Jahn-Teller impurity centers in a GaAs crystal [29] at a temperature of 25 K and in $E \times e$ $Ni^{2+}$ impurity impurity centers in a $Al_2O_3$ crystal at 2 K [14]. Note that the papers [14,28] actually discussed Raman's relaxation mechanism in tunneling. However, this mechanism was considered within the framework of the perturbation theory [21], which is not applicable to a weak inhomogeneous crystal field. The greates effect — the change in the average frequency of the spectrum of tunnel transitions due to two-phonon Raman processes — was not considered in these papers.

It also follows from Fig. 2 that $\delta$ and $\gamma$ decrease significantly when $|b - b_0|$ decreases and reach zero with $b = b_0$. The reason is that in the $E \times e$ case the quadratic vibronic interaction $b$ results in a trigonal distortion of the PES with the formation of three minima and three barriers with the height $bQ_0^2$. In the case of small $b$, the minima and the barriers are small, and the vibrations along and across the through (across and along the Jan-Teller distortions) are very different: the first vibrations are much softer than the latter. As $b$ increase, so do the barriers, and the vibrations along the through of PES become harder, but the vibrations across it soften. As a result, the difference in vibrations along and across the trough decreases with increase $b$. Accordingly, the mixing of phonons during the tunnel transition also decreases, which leads to a decrease in $\gamma$ and $\delta$. For $b = b_0 = 1/9$, each PES minimum became symmetric. At the same time, the spectra of vibrations in the minima have become equal and only the positions of the minima remain different. In this case, both $\gamma$ and $\delta$ are zero. With larger $b$, the difference in the phonon spectra appears again and begins to increase; at the same time, $\gamma$ and $\delta$ also increase. From Fig. 2 also follows that, there is a noticeable range for the values of the quadratic vibration interaction $b$, close to $b_0$, where the effect of this interaction on the tunnel transitions is actually turned off: in this range, the tunnel splitting begins to exceed $\delta \propto (T/T_D)^4$ and $\gamma \propto (T/T_D)^7$ already at a sufficiently high temperature. As a result, for such quadratic vibronic interactions, tunnel transitions remain coherent even at a sufficiently high temperatures.

## 6. Summary

In coherent tunnel transitions between different distorted states of impurity centers in crystals with the Jahn-Teller effect are considered, based on the assumption that the vibronic interaction with phonons is strong enough to destroy coherent tunneling. It has been established that Raman-type processes accompanying tunneling, caused by quadratic vibration interaction, lead not only to the broadening of the energy spectrum of transitions, but also to the deterioration of resonance. This greatly affects the temperature dependence of tunneling. At low temperatures $T \ll T_D$, the width and average frequency of the spectrum of tunnel transitions depend on the temperature according to the laws $\propto T^7/T_D{}^7$ and $\propto T^4/T_D{}^4$, respectively, where $T_D$ is the Debye temperature. The reason for the change in average frequency due to Raman processes is zero point vibrations, which make the creation of phonon in the Raman process more likely than the destruction of phonon with a similar frequency.

At low temperatures in crystals with weak inhomogeneous field, the rate of incoherent tunnel transitions changes with temperature depending on the inverse linear relationship: $\Omega^{-1} \propto T$. At high temperature, multiphonon processes become important, causing an increase in $\Omega$ with increasing temperature. Thus, the replacin the decline of $\Omega$ with an increase in temperature means the replacing the classic jumps over barrier with quantum tunneling through it. It should be noted that in experiments on ultrasound attenuation by $E \times e$ impurity centers of $Ni^{2+}$ ions in $Al_2O_3$ crystals, a change in the decrease to an increase in the rate of transition between different PES minima at a temperature 2 K was observed [14]. A similar effect was observed in the ultrasound attenuation due to tunnel transitions in the $T \times t$ Jahn-Teller centers in doped GaAs:Mn [28] and GaAs:Cu [29] crystals. However, the theoretical explanation for this change was not found in these works.

In the $E \times e$ case, it has been shown that there is a range of values for the quadratic vibration interaction close to the critical value of $b_0 = 1/9$ (in units of the Debye frequency), at which its effect on tunnel transitions disappears, and tunneling remains coherent at sufficiently high temperatures. The reason of existence of this range is that this interaction results not only in tunneling barriers, but also hardens vibrations along the trough of the PES and softens vibrations across the through. As a result, with a small quadratic vibronic interaction $b$, the mixing of phonons during the tunnel transition decreases with increasing $b$ (see Fig. 1). This abnormal relationship occurs to the critical value of $b = b_0$, when decrease is replaced by increase. For the critical value of $b$, phonon Hamiltonians in different PES minima have different equilibrium positions but the same spectra. Therefore, the quadratic vibronic interaction at $b \approx b_0$ does not lead to mixing of phonon during the tunnel transitions. As a result, Raman processes remain weak for $b \approx b_0$, and tunnel transitions remain coherent even at a relatively high temperatures.

## Appendix

To find the Fourier transform of the spectrum of an incoherent tunnel transition $F(t) = \left\langle e^{itH_{l'}} e^{-itH_l} \right\rangle$, we apply the shift operator $e^{\nabla_{l,l'}}$ to the evolution operators $e^{-itH_{1l}}$ and $e^{itH_{2l}}$. We get $F(t) = \left\langle e^{\nabla_{l'}} e^{it\tilde{H}_{l'}} e^{-it\tilde{H}_l} e^{\nabla_{ll'}(t)} e^{-\nabla_l} \right\rangle$, where $\nabla_{ll'} = \nabla_l - \nabla_{l'}$, $\nabla_{ll'}(t) = e^{itH_l} \nabla_{ll'} e^{-itH_l}$ is the time-dependent operator $\nabla_{ll'}$. At low temperatures, the greatest contribution to the rate of the tunnel transitions $\Omega$ comes from zero-phonon transitions with a small spectral width [23,24]. This contribution is given by $F(t)$ in the large $t$ limit. At this limit, operators decouple at very different times [24]. This gives $F(t)_{t\to\infty} = \left\langle e^{it\tilde{H}_{l'}} e^{-it\tilde{H}_l} \right\rangle \left\langle e^{\nabla_{ll'}} \right\rangle \left\langle e^{-\nabla_{ll'}} \right\rangle$. Using the Bloch-de Dominicis theorem, we obtain $\left\langle e^{\pm\nabla_{ll'}} \right\rangle_l = e^{-Z/2}$, where $Z = -\left\langle \nabla_{ll'}^2 \right\rangle$ is the Huang-Rhys (Ham) factor, which describes a decrease in the probability of zero-phonon transitions due to changes in the equilibrium positions of atoms during the transition. As a result, we get $F(t)_{t\to\infty} = e^{-Z} \tilde{F}(t)_{t\to\infty}$, where

$$\tilde{F}(t) = \left\langle e^{it\tilde{H}_{l'}} e^{-it\tilde{H}_l} \right\rangle \tag{A1}$$

is the Fourier transform in the case of purely quadratic vibronic interaction. To find $\tilde{F}(t)_{t\to\infty}$ in the case of multiple configurational coordinates involved in the quadratic vibronic interaction, we apply a method developed in [19,20] to compute vibronic transitions. This method is based on the use of the Dyson relation $e^{it(\tilde{H}_l+V)} \cong \hat{T} e^{it\sum_n V(t_n)} e^{it\tilde{H}_l}$ and the Stratonovich's identity. Here $t_n = nt/N$, $n = 1, 2, \ldots N \gg 1$, $\hat{T}$ arranges the operators from left to right according to the increasing $t_n$. We take $l = 1$ and $l' = 2$. Then $V = \bar{b}(Q_1^2 - \tilde{Q}_2^2)/2$, where $\bar{b} = b_2 - b_1$, $Q_1 = Q_{11}$, $\tilde{Q}_2 = (\sqrt{3}Q_{21} - Q_{11})/2$ ($Q_1$ and $\tilde{Q}_2$ are not orthogonal). Applying the Dyson relation now, we get

$$\tilde{F}(t) \cong \left\langle \hat{T}\, \mathrm{e}^{it(\bar{b}/2N)\sum_n (Q_1^2(t_n) - \tilde{Q}_2^2(t_n))} \right\rangle. \tag{A2}$$

Using now the Stratonovich's identity, we can present $\tilde{F}(t)$ as

$$\tilde{F}(t) \cong \left( \prod_n (2\pi)^{-1} \int\int du_{1n} du_{2n} e^{-(u_{1n}^2 + u_{2n}^2)/2} \right) \left\langle \hat{T} e^{\sqrt{it\bar{b}/N} \sum_n \left( Q_1(t_n) u_{1n} - \tilde{Q}_2(t_n) u_{2n} \right)} \right\rangle \tag{A3}$$

The quantum-statistical averaging can be performed using the Bloch–De Dominicis theorem. We get

$$\tilde{F}(t) = \prod_n \left( \frac{1}{2\pi} \int\int du_{1n} du_{2n} e^{-(u_{1n}^2 + u_{2n}^2)/2} \right) \exp\left[ \frac{-\bar{b}t}{8N} \sum_{nn'} \left( D_{nn'}^{(1)} u_{1n}(4u_{1n} + iu_{2n'}) - (D_{nn'}^{(1)} + 3D_{nn'}^{(2)}) u_{2n} u_{2n'} \right) \right].$$

Here $D_{nn'}^{(\iota)} \equiv D_\iota(t_n - t_{n'}) = i \left\langle \hat{T} Q_\iota(t_n) Q_\iota(t_{n'}) \right\rangle$ is the causal correlation function of coordinate $Q_\iota$. Now consider that the correlation function $D^{(i)}\left(t_n - t_{n'}\right)$ tends to zero if $|t_n - t_{n'}| \to \infty$ [24]. As a result, the matrix elements $D_{nn'}^{(i)}$ depend on the difference $n - n'$ in the $|t_n - t_{n'}| \to \infty$ limit.

Therefore, the eigenvectors and eigenvalues of the matrix $\hat{D}^{(i)}$ in this limit are equal to [19,20] $S_{kn} = N^{-1/2} e^{i\pi(2k+1)n/N}$ and $D_k^{(i)} = \sum_m e^{-i\omega_k m/N} D_i(tm/N)$, respectively, where $\omega_k = \pi(2k+1)/t$, $k = -N_0, -N_0+1, \ldots N_0$, $N = 2N_0$. Replacing variables $u_{in}$ by $x_{ik} = \sum_n S_{kn} u_{in}$, we get

$$\tilde{F}(t)_{t\to\infty} = \prod_k (2\pi)^{-1} \int\int dx_{1x} dx_{2k} \exp\left[-(x_{1k}^2 + x_{2k}^2)/2 + (\bar{b}t/2N)\sum_{ii'} D_{ii',k} x_{ik} x_{i'k}\right]. \quad \text{(A4)}$$

Here $D_{11,k} = D_k^{(1)}$, $D_{22,k} = -(D_k^{(1)} + 3D_k^{(2)})/4$, $D_{12,k} = D_{21,k} = D_k^{(1)}/4$. Integration over the variables $x_{1k}$ and $x_{2k}$ gives $\exp\left(-2^{-1}\ln\left|I + tbD_k/N\right|\right)$, where $|\ldots|$ is the determinant of the $2\times 2$ matrix. Having calculated this determinant, we obtain

$$\ln\tilde{F}(t)_{t\to\infty} = -\frac{1}{2}\sum_k \ln\left(\left(1 - \frac{\bar{b}tD_k^{(1)}}{N}\right)\left(1 + \frac{\bar{b}t(D_k^{(1)} + 3D_k^{(2)})}{4N}\right) + \left(\frac{\bar{b}tD_k^{(1)}}{4N}\right)^2\right). \quad \text{(A5)}$$

Let us replace the sums of $k$ and $m$ with integrals over $\omega$ and $t'$: $\sum_k \to \int dk = (t/2\pi)\int d\omega$, $\sum_m \to \int dm = (N/t)\int dt'$. We get

$$F(t)_{t\to\infty} = e^{-(\gamma+i\delta)t-Z}, \quad \text{(A6)}$$

where

$$\gamma - i\delta = \frac{1}{4\pi}\int d\omega \ln\left(\left(1 - \bar{b}D_1(\omega)\right)\left(1 + \frac{\bar{b}}{4}\left(D_1(\omega) + 3D_2(\omega)\right)\right) + \frac{\bar{b}^2}{4}D_1^2(\omega)\right) \quad \text{(A7)}$$

describes the spectral width ($\gamma$) and the average frequency of the tunnel transitions ($\delta$), $D_i(\omega) = \int dt e^{i\omega t} D_i(t)$ is the spectral representation of the correlation function $D^{(i)}(t)$. Simple calculations give

$$D_i(\omega) = G_i(\omega) + in_{|\omega|}\Gamma_i(|\omega|). \quad \text{(A8)}$$

Note that in the case of single configurational coordinate ($D_1(\omega) \neq 0$, $D_2(\omega) = 0$) equation (A7) coincides, up to notation, with Levenson's equation (28) in [32] and equation (12.80) in [34] obtained using the cumulant decomposition of the Fourier transform. At zero temperature $D_i(\omega)_{T=0} = G_i(\omega)$. Taking into account the relations $(1-\bar{b}G_1)(1+\bar{b}G_2) = 1$ and $(1-\bar{b}G_1)G_2 = G_1$, we get

$$\ln\left((1-\bar{b}G_1)\left((1+(\bar{b}/4)(G_1+3G_2)\right) + (\bar{b}^2/4)G_1^2\right) = 0 \quad \text{(A9)}$$

As a result, at zero temperature $\gamma$ and $\delta$ reaching zero, as it should be for physical reasons: the transition spectrum between the zero-point states of the energy-equivalent minimums of PES does not broaden or shift. At low temperatures $T \ll T_D$, keeping in Eq. (A7) the terms that are linear with respect to $n_\omega$, we get

$$\gamma \cong (3\bar{b}^2/4\pi)\int d\omega n_\omega \Gamma_1(\omega)\Gamma_2(\omega), \quad \text{(A10)}$$

$$\delta \cong 3(\bar{b}/\pi)\int d\omega n_\omega \left(\Gamma_1(\omega)\left(1 - \bar{b}\Delta_2/4\right) - \Gamma_2(\omega)\left(1 - \bar{b}\Delta_1/4\right)\right), \quad \text{(A11)}$$

For low frequency of phonons, $\Gamma_1 \cong 16\omega^3/(1-2b)^2$, $\Delta_1 \cong -2/(1-2b)$, $\Gamma_2 \cong 16\omega^3/(49b^2)$, $\Delta_2 \cong -2/(7b)$. By substituting these ratios into equations (A10) and (A11) and integrating them in frequency, we obtain equations (11) and (12).

**References**

1. H.A. Jahn and E. Teller, Proc. R. Soc. London, Ser. A, **161** 220 (1937).
2. I.B. Bersuker, Sov. Phys. JETP **16** 933 (1963).
3. Mary C. O'Brien, Proc. R. Soc. A. **281** (1384) 323 (1964).
4. F.S. Ham,.Phys. Rev. **138** A1727 (1965).
5. R. Englman, "The Jahn-Teller Effect in Molecules and Crystals", Wiley London (1972).
6. C.A. Bates "Jahn–Teller effects in paramagnetic crystals". Physics Reports **35** 187 (1978).
7. M.D. Sturge, The Jahn-Teller efferct in Solids, Solid State Physics **20** 91 (1968).
8. P. Garcia-Fernandez, A. Trueba, M.T. Barrius, J.A. Aramburu, Miguel Mareno, "Dynamic and Static Jahn-Teller Effect in Impurities: Determination of Tunnel Splitting", in: Progress in Theoretical Chemistry and Physics (PTCP) **23** Springer Dordrecht (2011).
9. I. B. Bersuker and V. Z. Polinger, "Vibronic interactions in Molecules and Crystals", Springer-Verlag Berlin-Heidelberg (1989).
10. I.B. Bersuker, "The Jahn-Teller Effect", Cambridge University Press, Cambridge (UK) (2006).
11. Yu.E. Perlin, B.S. Tsukerblat, "Electron-Vibrational Interaction in Optical Spectra of Impurity Paramagnetic ions", Stinicca, Kishinev (1974) (in Russian).
12. Y. Toyozawa, "Optical processes in Solids", Cambridge University Press (2003).
13. V. Hizhnyakov, H. Kaasik, and I Sildos, Phys. Stat. Sol. **234** 644 (2002)); V. Hizhnyakov, V. Boltrushko, H. Kaasik and I. Sildos, J. Chem. Phys. **119** 6290 (2003).
14. M.D. Sturge, J.T. Krause, E.M. Gyorgy, R.C. LeCraw, F.R. Merrit, Phys. Rev. **155** 218 (1967).
15. H. Koizumi, I. Bersuker, Phys. Rev. Lett. **83** 3009 (2000).
16. Kaja Pae, V. Hizhnyakov, J. Chem. Phys. **147** 084107 (2017).
17. Kaja Pae, V. Hizhnyakov, J. Chem. Phys. **141** 234113 (2014).
18. Kaja Pae, V. Hizhnyakov, J. Chem. Phys **145** 064108 (2016).
19. V. Hizhnyakov, Imbi Tehver, Chem. Phys. Lett. **422** 299 (2006).
20. V. Hiznyakov, Chem. Phys. Lett. **493** 19 (2010); **808** 140092 (2022).
21. K. Suzuki and N. Mikava, Phys. Rev. Lett 28 96 (1972).
22. D.E. McCumber and M.D. Surge, J. Appl. Phys. **34** 1682 (1963).
23. D.E. McCumber Math. Phys. **5** 508 (1964).
24. M.A. Krivoglaz, Zh. Eksp. Teor. Fiz. **46** 637 (1964).
25. K. Rebane, "Impurity Spectra of Solids", Plenum Press, New York, 1970.
26. O. Sild, K. Haller (Eds.), "Zero-Phonon Lines", Springer Berlin Heidelberg (1988).
27. G. I. Bersuker and V. Z. Polinger, Zh. Eksp. Teor. Fiz. 80 1798 (1981).
28. K Lassmann and Hp. Schad, Solid State Commun. **44** 449 (1976).
29. N.S. Averkiev, K.A. Baryshnikov, I.B. Bersuker, V.V. Gudkov, I.V. Zhevstovskikh, V.Yu. Mayakin, A.M. Monakhov, M.N. Sarychev,V.E. Sedov, JETP Letters **96** 236 (2012).
30. A. A. Maradudin, E. W. Montroll and G. H. Weiss, Theory of Lattice Dynamics in the Harmonic Approximation, Academic, New York, 1963.
31. M. Lax, J. Chem. Phys. **20** 1752 (1952).
32. G.F. Levenson, phys. stat. Sol. **43** 739 (1971).
33. I.S. Osad'ko, Sov. Phys. Solid State **13** 974 (1971); **17** 2098 (1975).
34. I.S.Osad'ko, "Selective Spectroscopy of Single Molecules", Springere (2003).
35. D. Hsu, J.L. Skinner, J. Lumines. **37**, 331 (1987).
36. R.L Stratonovich, Doklady Akademii Nauk 115 10 (1957). )W.P. Mason and T. B. Bateman, Phys. Rev 134 A1387 (1964) W.P. Mason and T. B. Bateman, Phys. Rev 134 A1387 (1964)